\documentclass[fleqn,usenatbib,twocolumn]{aastex63}

\received{XXX}
\revised{YYY}
\accepted{ZZZ}
\submitjournal{ApJL}

\shorttitle{Maximum NS mass and GW190814}

\usepackage{newtxtext,newtxmath}

\usepackage[T1]{fontenc}

\DeclareRobustCommand{\VAN}[3]{#2}
\let\VANthebibliography\thebibliography
\def\thebibliography{\DeclareRobustCommand{\VAN}[3]{##3}\VANthebibliography}

\usepackage{graphicx}	
\usepackage{amsmath}	
\usepackage{amssymb}	
\usepackage{color}

\def\Msun{{\mathrm{M}_{\odot}}}
\def\Mmax{{M_{\max}}}

\graphicspath{{./}{figures/}}

\begin{document}

\title[Maximum NS mass and GW190814]{On the maximum mass of neutron stars and GW190814}

\author[0000-0001-7454-0104]{Daniel A. Godzieba}%
\email{dag5611@psu.edu}
\affiliation{Department of Physics, The Pennsylvania
State University, University Park, Pennsylvania 16802}

\author[0000-0001-6982-1008]{David Radice}
\email{dur566@psu.edu}
\affiliation{Institute for Gravitation \& the Cosmos, The Pennsylvania State University, University Park, PA 16802}
\affiliation{Department of Physics, The Pennsylvania
State University, University Park, Pennsylvania 16802}
\affiliation{Department of Astronomy \& Astrophysics, The Pennsylvania State University, University Park, PA 16802}
 
\author[0000-0002-2334-0935]{Sebastiano Bernuzzi}
\email{sebastiano.bernuzzi@uni-jena.de}
\affiliation{Theoretisch-Physikalisches Institut, Friedrich-Schiller-Universit{\"a}t Jena, 07743, Jena, Germany}

\begin{abstract}
Motivated by the recent discovery of a compact object with mass in the range $2.5{-}2.67\, \mathrm{M}_\odot$ in the binary merger GW190814, we revisit the question of the maximum mass of neutron stars (NSs). We use a Markov Chain Monte Carlo approach to generate about 2 million phenomenological equations of state with and without first order phase transitions. We fix the crust equation of state and only assume causality at higher densities. We show how a strict upper bound on the maximum NS mass can be inferred from upcoming observation of NS radii and masses. The derived upper bounds depend only on relativity and causality, so it is not affected by nuclear physics uncertainties. We show how a lower limit on the maximum mass of NSs, in combination with upcoming measurements of NS radii by LIGO/Virgo and NICER, would constrain the equation of state of dense matter. Finally, we discuss the implications for GW190814.
\end{abstract}




\section{Introduction}
One of the most striking predictions of general relativity is the existence of a maximum mass for any static matter configuration. The value of the maximum mass for nonrotating neutron stars (NSs; $\Mmax$) is known to be the most important parameter controlling the outcome of binary NS mergers \citep{Shibata:2006nm, Sekiguchi:2011zd, Hotokezaka:2013iia, Bauswein:2013jpa, Palenzuela:2015dqa, Bernuzzi:2015opx, Lehner:2016lxy, Dietrich:2016hky, Piro:2017zec, Radice:2017lry, Koppel:2019pys}, and determines the possible formation of black holes (BHs) in core-collapse supernovae \citep{OConnor:2010moj, Schneider:2020kxr}. Moreover, the knowledge of the maximum NS mass would strongly constrain the (poorly known) equation of state (EOS) of matter at several times nuclear density \citep{Hebeler:2010jx, Lattimer:2012nd, Hebeler:2013nza, Ozel:2016oaf, Annala:2017llu, Tews:2018iwm, Annala:2019puf, Miller:2019nzo}.

Recently, the LIGO/Virgo collaboration announced the discovery of a compact binary merger, GW190814, containing a compact object with mass $2.5{-}2.67\, \Msun$ \citep{Abbott:2020khf}. This is either the most massive NS, or the least massive BH ever found. Gravitational wave (GW) observations did not reveal the nature of this object. However, understanding whether it was a NS or a BH would have profound implications for our understanding of high-density physics and the formation of compact objects in the Universe \citep{Abbott:2020khf}.

The theoretical upper limit on $\Mmax$ is of $3.2\, \Msun$ \citep{Rhoades:1974fn}. This limit rests on weak assumptions on the nature of nuclear forces at around nuclear saturation density ($\rho_\text{nuc} \simeq 2.7 \times 10^{14}\, {\rm g}\, {\rm cm}^{-3}$), and on causality. It can be somewhat reduced if the EOS of matter is assumed to be known (within some uncertainty range) up to some given density \citep[e.g.,][]{Kalogera:1996ci, Hebeler:2013nza, Tews:2019qhd}. Observationally, only lower bounds on the maximum mass of nonrotating NSs are known with high confidence from mass measurements of slowly rotating NSs. The precise determination of the mass of pulsars J1614-2230 ($1.908 \pm 0.016\, \Msun$; \citealt{Demorest:2010bx, Fonseca:2016tux, Arzoumanian:2017puf}), J0348+0432 ($2.01 \pm 0.04\, \Msun$; \citealt{Antoniadis:2013pzd}), J0740+6620 ($2.14 \pm 0.1\, \Msun$; \citealt{Cromartie:2019kug}), and J2215+5135 ($2.27 \pm 0.17\, \Msun$; \citealt{Linares:2018ppq}) show that $\Mmax$ should be of at least $2\, \Msun$. Multimessenger observations of the binary NS merger GW170817 \citep{GBM:2017lvd} have been interpreted as an indication of an upper bound on $\Mmax$ of about $2.3\, \Msun$ \citep{Margalit:2017dij, Rezzolla:2017aly, Shibata:2019ctb, Ruiz:2017due, Shibata:2017xdx}. However, this interpretation rests on the poor understanding of the long-term postmerger evolution of NS binaries \citep{Radice:2018xqa, Radice:2020ddv} and is not universally accepted \citep{Ai:2018jtv, Li:2018hzy, Piro:2018bpl}. The tidal deformability data from GW170817 also suggests that $\Mmax$ should be of about $2.3\, \Msun$ \citep{Lim:2018bkq, Abbott:2018exr, Essick:2019ldf}, but these results are very sensitive to their prior and model choices \citep{Greif:2018njt, Lim:2020zvx}. This is not surprising, since GW170817 does not probe directly the EOS at the densities relevant for NSs close to the maximum mass. Finally, the statistical analysis of the mass distribution of known NSs \citep{Alsing:2017bbc, 2020RNAAS...4...65F} also provides a plausible range for $\Mmax$, but this analysis suffers from large uncertainties due to the small number of known NSs with high mass and due to possible selection effects. Summarizing, $\Mmax$ is currently only weakly constrained. 

The announcement of GW190814 triggered a renewed interest in the possibility of very massive NSs and the possible implications of the their existence. \citet{Most:2020bba}, \citet{Zhang:2020zsc}, and \cite{Dexheimer:2020rlp} argued that the presence of a NS in GW190814 is compatible with a relatively small $\Mmax$ under the assumption that the NS was rapidly spinning. However, such scenario requires significantly faster rotation than that of any observed millisecond pulsar. Indeed, in order to significantly exceed the maximum mass, a rotating NS must be endowed with a few times $10^{52}\, \mathrm{erg}$ in rotational kinetic energy \citep{Margalit:2017dij}\footnote{See, however, \citet{Safarzadeh:2020ntc} for a proposed mechanism for the formation of fast spinning compact objects in binaries.}. More exotic interpretations include anisotropic NSs \citep{Roupas:2020jyv} and primordial BHs \citep{Vattis:2020iuz}. \citet{Tsokaros:2020hli} and \citet{Fattoyev:2020cws} showed that current EOS models can accommodate present astrophysical constraints on the NS radii and, at the same time, explain a $2.6\, \Msun$ NS, although the resulting EOS is in tension with constraints from heavy-ion collision experiments \citep{Danielewicz:2002pu}.  Finally, the possible implications of the existence of a $2.6\, \Msun$ NS for the EOS of dense matter have been explored by \citet{Tews:2020ylw} and \citet{Lim:2020zvx}. 

Despite the intense theoretical efforts, the nature of the secondary in GW190814 remains unknown. Here, we derive strict upper bounds on $\Mmax$ under the sole assumptions of causality and general relativity. With these assumptions, only bulk properties of NSs factor into the upper bound, and thus the upper bound is independent of nuclear physics uncertainties. We show that current astrophysical constraints on the NS EOS are not in tension with a NS in GW190814, but that future measurements of radii and tidal deformabilities could translate into more stringent upper bounds on $\Mmax$ and exclude that GW190814 was a NS-BH, unless extreme rotation is invoked for the secondary object.

\section{Methods}
The constraints of causality and general relativity define a space of all possible EOSs, referred to as the EOS band. To probe this space in a computationally efficient manner, we utilize a Markov chain Monte Carlo (MCMC) algorithm. We parameterize the EOSs within the band using a variation on the method developed by  \citet{Read:2008iy} that approximates each EOS as a continuous piecewise polytrope with four pieces: \begin{equation}
    p(\rho) = \begin{cases} K_0 \rho^{\Gamma_0} \quad \rho \leq \rho_0 \\ K_1 \rho^{\Gamma_1} \quad \rho_0 < \rho \leq \rho_1 \\ K_2 \rho^{\Gamma_2} \quad \rho_1 < \rho \leq \rho_2 \\ K_3 \rho^{\Gamma_3} \quad \rho > \rho_2. \end{cases}
    \label{EOS}
\end{equation} The choice of a piecewise polytropic ansatz does not significantly bias the results compared to other EOS representation schemes \citep{Annala:2019puf,Annala:2017llu}. The first polytrope piece corresponds to the (presumed known) crust EOS, with $K_0 = 3.59389\times10^{13}$~[cgs] and $\Gamma_0 = 1.35692$ \citep{Douchin:2001sv}, and is fixed for all EOSs. The choice of the crust EOS does not influence the bulk properties of massive NSs \citep{Rhoades:1974fn, Most:2018hfd}. The last three pieces are specified by six parameters: three transition densities, $\rho_0 \in [0.15\rho_\text{nuc},1.2\rho_\text{nuc}]$, $\rho_1 \in [1.5\rho_\text{nuc},8\rho_\text{nuc}]$, $\rho_2 \in [\rho_1 ,8.5\rho_\text{nuc}]$; and three adiabatic indices, $\Gamma_1 \in [1.4,5]$, $\Gamma_2 \in [0,8]$, $\Gamma_3 \in [0.5,8]$. Nuclear saturation density is taken to be $\rho_{\rm nuc} = 2.7 \times 10^{14}\, {\rm g}\, {\rm cm}^{-3}$. The polytropic constants $K_1$, $K_2$, and $K_3$ are fixed by requiring the continuity of the EOS. The wide bounds on the possible values for the $\Gamma_i$ allow for a diverse variety of EOSs with a wide range of softnesses and includes EOSs with and without first-order phase transitions.

The MCMC algorithm will be discussed in detail in a future publication (Godzieba et al., in prep). The basic aspects of the algorithm are as follows. For each trial EOS we compute sequences of solutions of the  Tolman-Oppenheimer-Volkoff (TOV) equations using the publicly available \texttt{TOVL} code that is described in \citet{Bernuzzi:2008fu} and \citet{Damour:2009vw}. The transition probability is determined by whether the physical properties of the EOS are within three weak physical constraints: 1) causality of the maximum mass NS (sound speed $c_s < c$); 2) $\Mmax > 1.97 \, \Msun$; 3) tidal deformability of the $1.4\, \Msun$ NS $\Lambda_{1.4} < 800$. We have verified that our results do not change if the upper limit on $\Lambda_{1.4}$ is set to 4,000. Here, we report the results from the analysis performed with $\Lambda_{1.4} < 800$, since this allows for a more dense coverage of the relevant portion of the parameter space.

\section{Results}
We assemble a data set of 1,966,225 phenomenological EOSs. As already mentioned, the maximum value of $\Mmax$ in our data set is $2.9\, \Msun$. Much larger masses are found if $\Lambda_{1.4}$ is allowed to be larger than $800$, however these EOSs are strongly disfavored in light of GW170817 \citep{Abbott:2018wiz, LIGOScientific:2018mvr}. In any case, our conclusions below would not be altered if we included these more stiff EOSs. Our analysis reveals that the set of all EOSs satisfying certain conditions on the NS radii admits stricter upper bounds on $\Mmax$. To be concrete, we consider cases in which the radii of reference NSs with masses $1.4\, \Msun$ and $2.14\, \Msun$ -- $R_{1.4}$ and $R_{2.14}$ -- are fixed, and we show that more stringent upper bounds on $\Mmax$ can be derived in these cases.

We consider at first the impact of restricting the range of $R_{1.4}$. The radius of a NS with mass close to $1.4\, \Msun$, PSR J0030+0451, has been directly measured by NICER \citep{Riley:2019yda, Miller:2019cac}. An indirect constraint on $R_{1.4}$ has also been obtained using multimessenger data from GW170817 \citep{Annala:2017llu, Most:2018hfd, De:2018uhw, Abbott:2018exr, Radice:2018ozg, Capano:2019eae, Essick:2019ldf}. More precise constraints are expected as systematics in NICER data are better understood, and when GW observatories will come back online with increased sensitivity. Finally, $R_{1.4}$ is known to correlate strongly with the EOS at around twice nuclear saturation density \citep{Lattimer:2012nd}. This makes $R_{1.4}$ a particularly interesting case.

\begin{figure}
    \centering
    \includegraphics[width=\columnwidth]{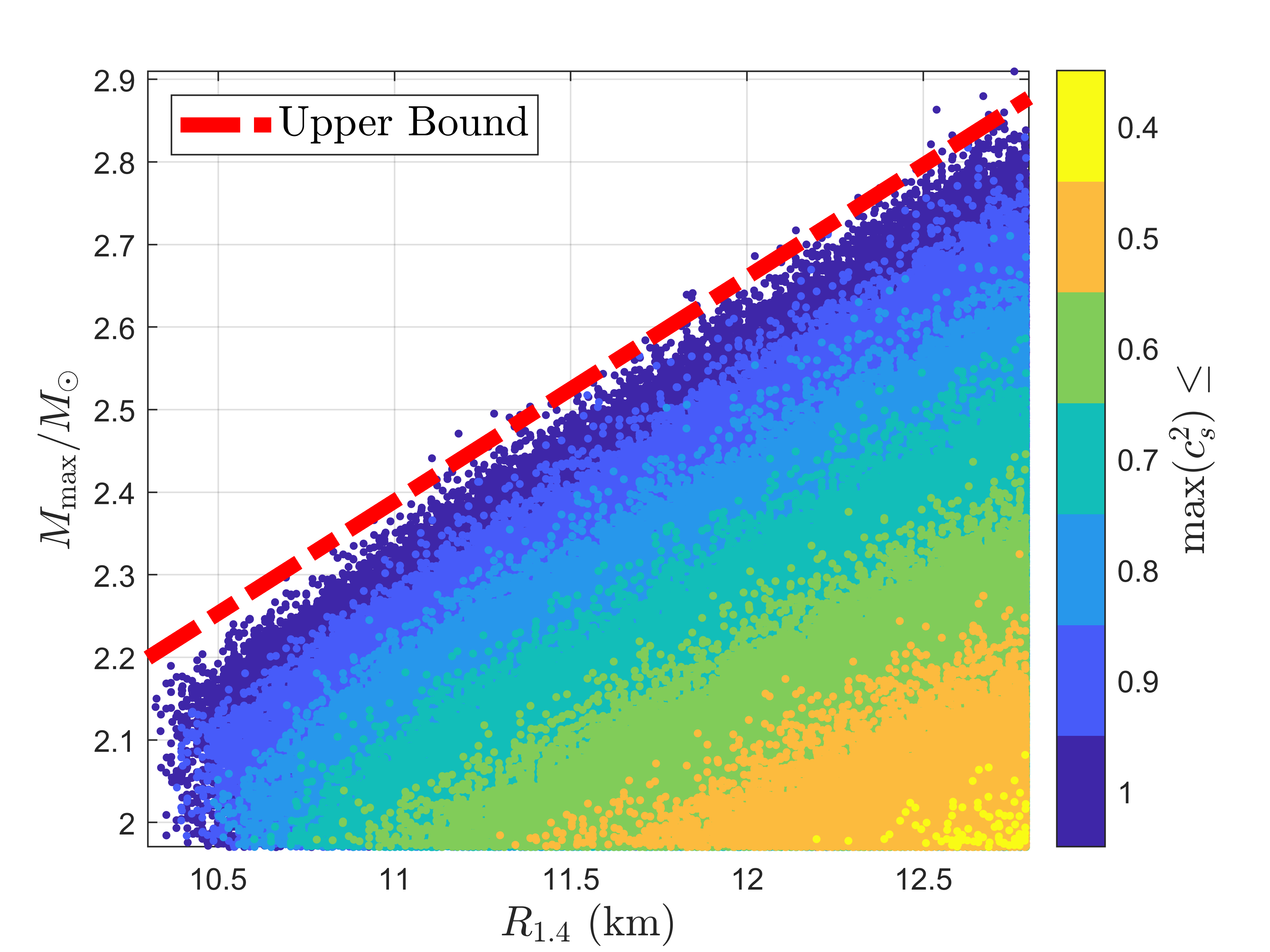}
    \caption{Distribution of $\Mmax$ and radius of the $1.4\, \Msun$ NS, $R_{1.4}$ for about 2 million phenomenological EOSs. Each point is colored according to the value of the sound speed reached in the maximum mass NS (points with smaller $\max(c_{s}^{2})$ are drawn on top). The red dashed line shows an approximate linear ansatz for the boundary of all physical EOSs in the $\Mmax-R_{1.4}$ plane (see main text for the details).}
    \label{fig:Mmax.R14}
\end{figure}

Fig. \ref{fig:Mmax.R14} shows the joint span of $\Mmax$ and $R_{1.4}$ across our data set. We also color each data point according to the maximum value of the sound speed reached in the maximum mass NS predicted by that EOS. As expected in the light of the results of \citet{Rhoades:1974fn}, we find that the largest $\Mmax$ values are reached at the boundary of causality, that is when $c_s = c$ at the highest densities. We also find that the range of $\Mmax$ decreases substantially as $R_{1.4}$ decreases. For example, an upper limit on $R_{1.4}$ of $11\, {\rm km}$ would imply $\Mmax \lesssim 2.35\, \Msun$. NICER observations of PSR J0030+0451 currently only provide a weak upper bound of $R_{1.4} \lesssim 14\, {\rm km}$ \citep{Miller:2019cac}. This measurement is currently not constraining for $\Mmax$. GW and electromagnetic (EM) observations of the NS merger in GW170817 place a more stringent constraint $R_{1.4} \lesssim 13\, {\rm km}$ \citep{De:2018uhw, Abbott:2018exr, Radice:2018ozg}. However, even this value is still compatible with maximum NS masses of up to ${\sim}2.9\, \Msun$. This is consistent with the findings of \citet{Tews:2019qhd} and \citet{Fattoyev:2020cws}, who constructed EOS models compatible with GW170817 and reaching maximum masses of ${\sim}2.6-2.9\, \Msun$.

\begin{figure}
    \centering
    \includegraphics[width=\columnwidth]{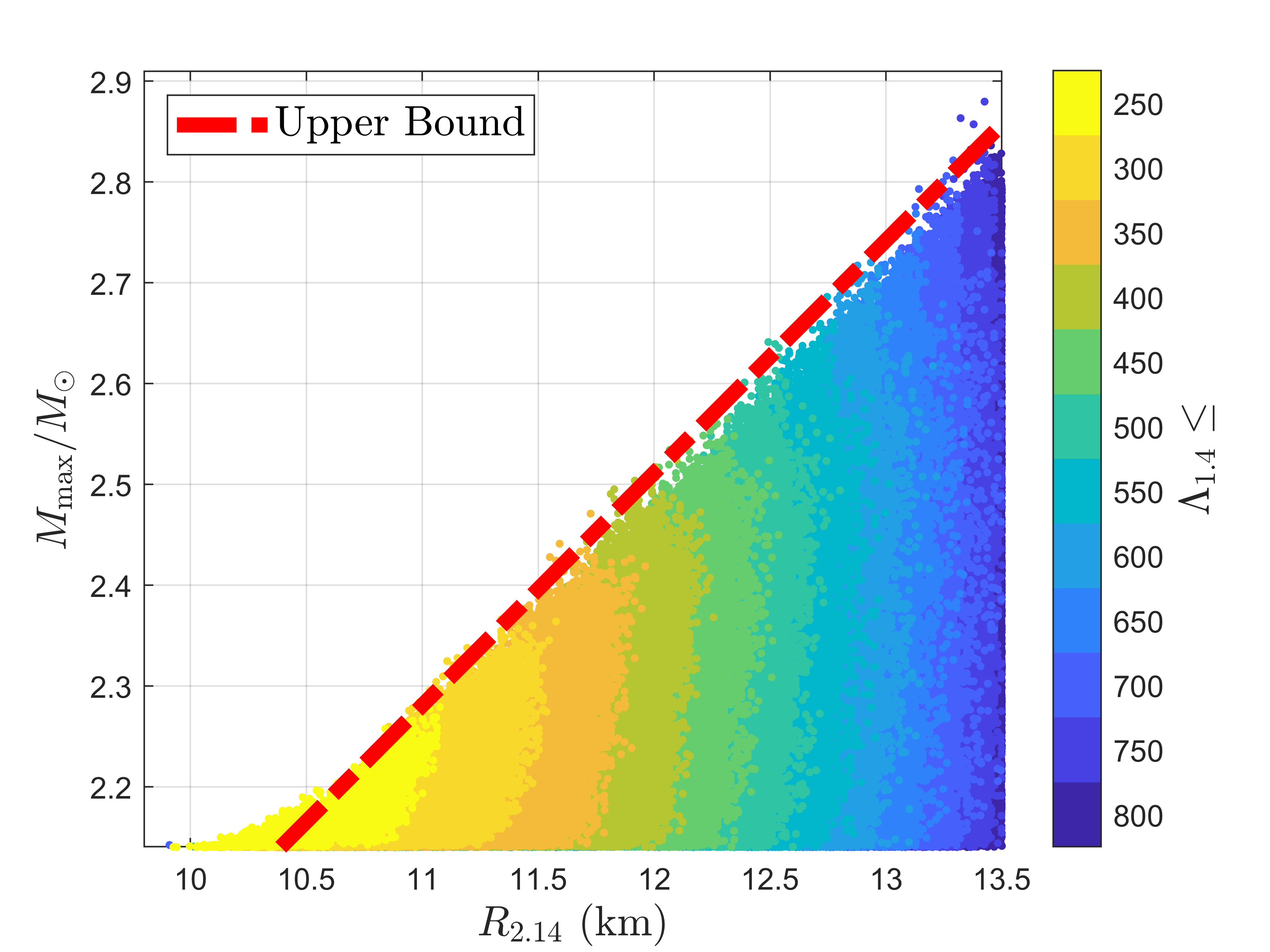}
    \caption{Distribution of $\Mmax$ and radius of the $2.14\, \Msun$ NS, $R_{2.14}$ for about 2 million phenomenological EOSs. Each point is colored according to the tidal deformability of the $1.4\,\Msun$ NS, $\Lambda_{1.4}$ (points with smaller $\Lambda_{1.4}$ are drawn on top). The red dashed line shows an approximate linear ansatz for the boundary of all physical EOSs in the $\Mmax-R_{2.14}$ plane (see main text for the details).}
    \label{fig:Mmax.R214}
\end{figure}

Among the next NICER targets are pulsars J1614-223 ($M \simeq 1.908\, \Msun$) and J0740+6620 ($M \simeq 2.14\, \Msun$). The measurement of their radii has the potential to yield very strong constraints on the EOS of dense matter \citep[e.g.,][]{Han:2020adu}, but also to constrain $\Mmax$, as we show in Fig.~\ref{fig:Mmax.R214}. We find that an upper bound on $R_{2.14}$ of $12\, {\rm km}$ would be sufficient to confidently rule out a NS in GW190814, unless fast rotation is invoked. Similarly, if $\Lambda_{1.4}$ can be constrained to be less than $400$ with future GW observations of merging NSs, then $\Mmax$ would be constrained to be less than ${\sim}2.5\, \Msun$.

In general, we find that the upper bound on $\Mmax$ can be very well approximated as \begin{equation}
    \Mmax \leq \alpha(M) + \beta(M) R_M, 
\end{equation} where $R_M$ is the radius in km of a NS of gravitational mass $M$ and \begin{equation}
    \begin{aligned}
        \alpha &= 0.45\, \Msun - 1.22\, M, \\
        \beta  &= -0.051\, \Msun\, {\rm km}^{-1} + 0.34\, M\, {\rm km}^{-1}. 
    \end{aligned}
\end{equation} These coefficients are obtained by performing a linear fit to a set of data points at the upper edge of the distribution that fall within the range $0.95 < \max(c_s^2) < 1$ for eight values of $M$. We then add a small shift of $0.04\, \Msun$ to $\alpha$, which is sufficient to enclose the vast majority of data points. The quality of these approximate expressions can be appreciated from Figs.~\ref{fig:Mmax.R14} and \ref{fig:Mmax.R214}.

\begin{figure}
    \centering
    \includegraphics[width=\columnwidth]{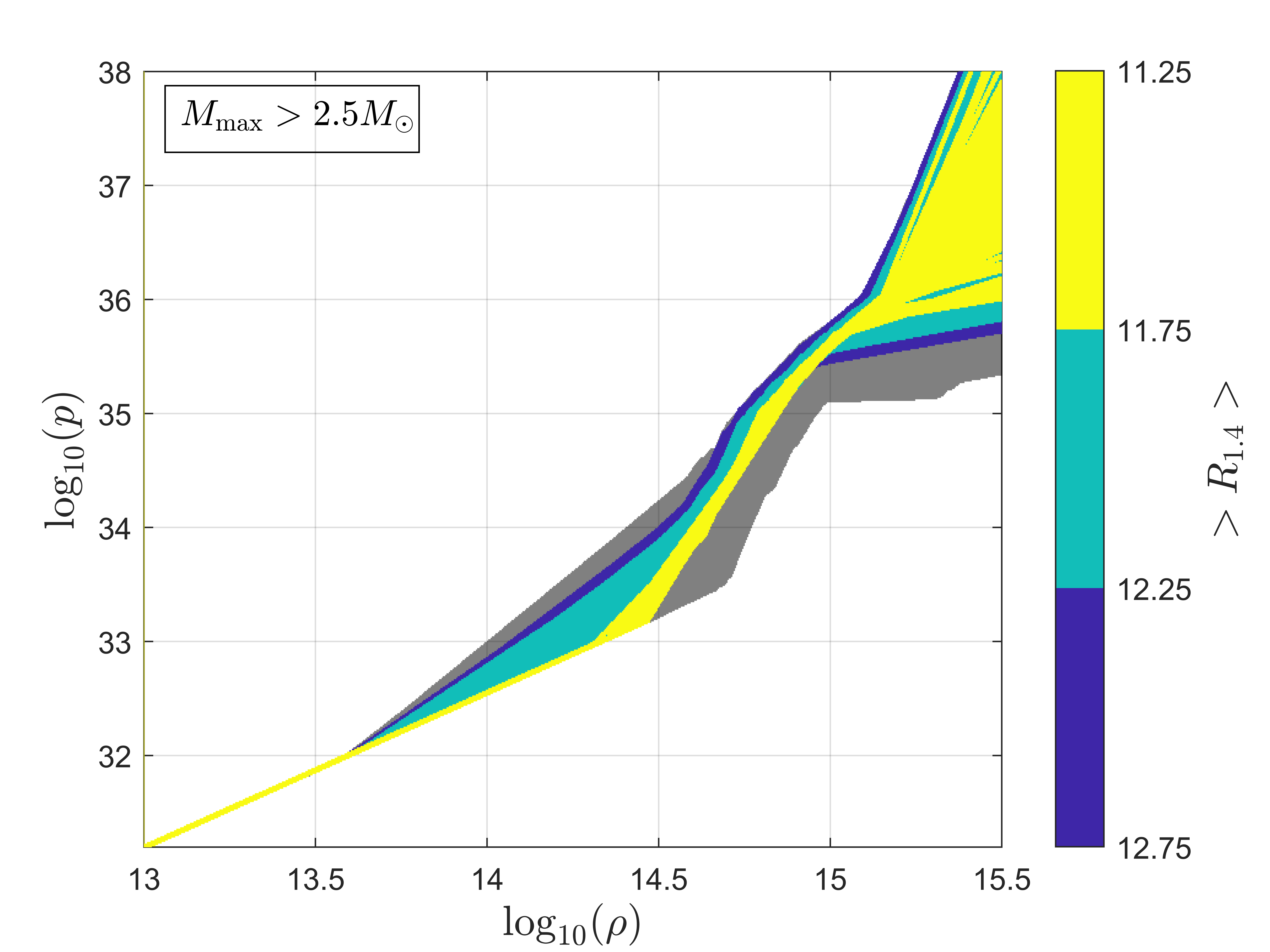}
    \caption{All EOSs with $\Mmax > 2.5 \Msun$ and $11.25 \ {\rm km} < R_{1.4} < 12.75 \ {\rm km}$ plotted as the pressure, $p$ ($\rm dyne/cm^2$), as a function of density, $\rho$ ($\rm g/cm^3$).  EOSs with smaller radii are drawn on top, since they cover a smaller region. The grey shaded region beneath the colored layers represents the full data set of 1,966,225 EOSs.}
    \label{fig:EOS band R_1.4}
\end{figure}

\begin{figure}
    \centering
    \includegraphics[width=\columnwidth]{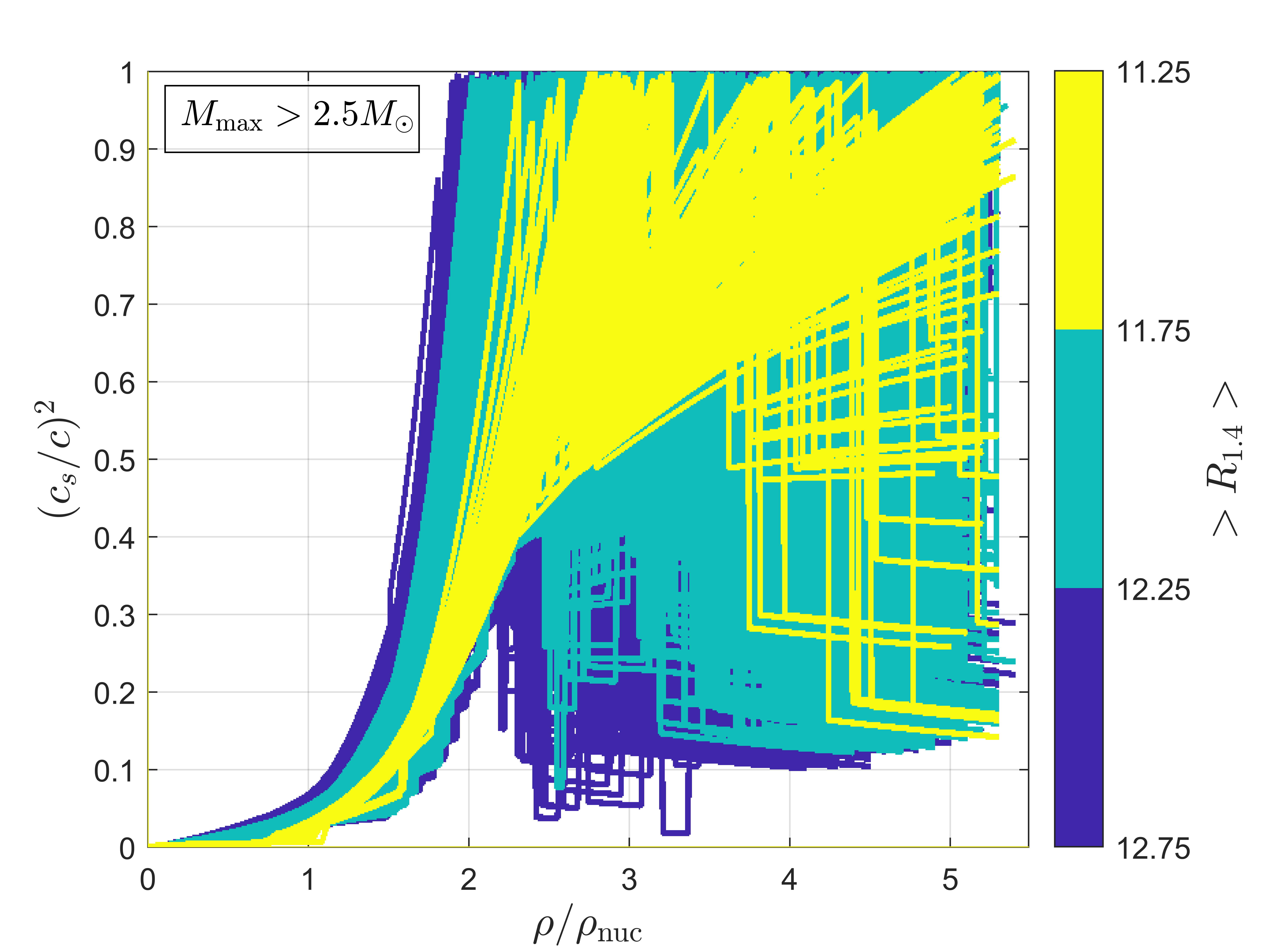}
    \caption{Sound speed squared for the EOSs with $\Mmax > 2.5 \Msun$ and $11.25 \ {\rm km} < R_{1.4} < 12.75 \ {\rm km}$ as a function of density, $\rho$ (normalized to $\rho_{\rm nuc}$). EOSs with smaller radii are drawn on top, since they cover a smaller region. We remark that the sound speed is not assumed to be smooth in our EOSs.}
    \label{fig:cs2 rho R_1.4}
\end{figure}

If the secondary in GW190814 is a NS, then the lower bound on the maximum mass is $\Mmax > 2.5\, \Msun$. Under this assumption, the measurement of the radius of the $1.4\, \Msun$ NS would strongly constrain the EOS up to densities of ${\sim}3 \rho_{\rm nuc}$ ($\log \rho \simeq 14.9$), as shown in Fig.~\ref{fig:EOS band R_1.4}. We find that an EOS supporting $\Mmax > 2.5\, \Msun$ would violate the causal limit if $R_{1.4} \lesssim 11.38\ {\rm km}$, which can be roughly seen in Fig.~\ref{fig:Mmax.R14}. We also find that all EOSs supporting $\Mmax > 2.5\, \Msun$ with radii in the range we considered for $R_{1.4}$ must violate the conformal limit of $(c_s/c)^2 < 1/3$ at ${\sim}2 \rho_{\rm nuc}$ (see Fig.~\ref{fig:cs2 rho R_1.4}). The behavior of the sound speed at higher densities depends on $R_{1.4}$. Among these constrained EOSs, those that accommodate small radii (between $11.38\, {\rm km}$ and $11.75\, {\rm km}$) must have $\max(c_{s}^{2}) \gtrsim 0.86$. So, while there are extreme EOS very close to the causal limit that lie in the small radius range, there are still EOS in this range that do not approach the causal limit. Thus, we cannot constrain out non-extreme EOS with $\Mmax > 2.5\, \Msun$ and $R_{1.4} < 11.75\ {\rm km}$. (If we look at EOSs that support $\Mmax > 2.6\, \Msun$, we find that the causal limit is violated if $R_{1.4} \lesssim 11.8\, {\rm km}$, which excludes the previous small radius range.) These measurements would still leave the extreme density part of the EOS ($\rho > 10^{15}\, {\rm g}\, {\rm cm}^{-3}$) relatively unconstrained. This would likely change if we were to enforce a matching to perturbative QCD at high densities \citep{Annala:2019puf}. We leave this analysis to future work.

\section{Conclusions}
We revisited the problem of determining the maximum mass for nonrotating NSs $\Mmax$ using minimal assumptions and a nuclear-physics agnostic approach. Differently from other works, our analysis does not attempt to provide a likely value $\Mmax$ and is not affected by prior or model choices. Instead, we provide strict upper limits based on the sole assumptions of causality and general relativity. Our results confirm that current astrophysical measurements of radii and tidal deformability of NSs with canonical mass ${\sim}1.4\ \Msun$ do not significantly constrain $\Mmax$. In particular, we find that the presence of a nonrotating or moderately spinning NSs in GW190814 cannot be excluded. However, future observations constraining the radii of massive NSs have the potential to yield stringent upper bounds on the NS maximum mass and might rule out a nonrotating NSs in GW190814. Our analysis also show that, if the secondary in GW190814 was in fact a NS, then this information, in combination with upcoming measurements of $R_{1.4}$, would strongly constraint the EOS to densities up to ${\sim}3 \rho_{\rm nuc}$. It would also imply that the conformal limit must be violated at ${\sim}2\rho_{\rm nuc}$. However, even a stringent upper limit on $R_{1.4}$ of $11.75\, {\rm km}$, together with $\Mmax > 2.5\, \Msun$, could still be accommodated by EOSs that are well below the causal limit. If we assume $\Mmax > 2.5\, \Msun$, the causal limit is strictly violated for $R_{1.4} \lesssim 11.38\, {\rm km}$, and if we assume $\Mmax > 2.6\, \Msun$, the causal limit is strictly violated for $R_{1.4} \lesssim 11.8\, {\rm km}$.

Our analysis did not make use of the knowledge of the EOS of matter at around nuclear density \citep{Gandolfi:2019zpj}. Instead, we treated the EOS at densities beyond those of the crust as being unconstrained. It is likely that, with the inclusion of more information from nuclear theory, more stringent upper bounds on $\Mmax$ could be derived. However, the fact that our results are broadly consistent with those obtained with more sophisticated approaches \citep{Hebeler:2013nza, Annala:2017llu, Tews:2019qhd, Fattoyev:2020cws} suggests that the extent to which the upper bound might be improved will be modest. Indeed, it is well known that $\Mmax$ depends most strongly on the EOS at densities of several times that of nuclear saturation \citep{Lattimer:2012nd}. 

Our analysis did not consider the possibility of strange quark stars, but our EOS parametrization does allow for hybrid stars with hadronic crusts and a first order QCD phase transition in their interior \citep[e.g.,][]{Annala:2019puf}. We leave the determination of upper bounds on the maximum mass for self-bound quark stars to a future work. Future work should also consider the implication of X-ray burst \citep{Steiner:2010fz, Lattimer:2012nd, Ozel:2016oaf} and laboratory constraints \citep{Danielewicz:2002pu} on the high density EOS on the upper bound of $\Mmax$.

\section*{Software used}
TOVL \citep{Bernuzzi:2008fu}, Matlab\textsuperscript{\textregistered}.

\section*{Data Availability}
EOS parameters and bulk properties of reference NSs generated for this work are publicly available on Zenodo \citep{godzieba_daniel_a_2020_3954899}.

\section*{Acknowledgements}
We thank the anonymous referee for insightful suggestions that improved the quality of this work.
Computations for this research were performed on the Pennsylvania State University's Institute for Computational and Data Sciences Advanced CyberInfrastructure (ICDS-ACI).
S.B. acknowledges support by the EU H2020 under ERC Starting Grant, no.~BinGraSp-714626.

\bibliography{NSMaxMass}{}
\bibliographystyle{aasjournal}



\end{document}